\documentclass[twocolumn,english,superscriptaddress,prl,amsmath,amssymb,floatfix,longbibliography]{revtex4-1}
\usepackage[T1]{fontenc}
\usepackage[utf8]{inputenc}
\usepackage{color}
\usepackage{babel}
\usepackage{verbatim}
\usepackage{bm}
\usepackage{amsmath}
\usepackage{graphicx}
\usepackage{esint}
\usepackage[unicode=true,pdfusetitle,
 bookmarks=true,bookmarksnumbered=false,bookmarksopen=false,
 breaklinks=false,pdfborder={0 0 1},backref=false,colorlinks=true]
 {hyperref}
\hypersetup{
 pdfborderstyle=}

\makeatletter
\usepackage{babel}

\usepackage{dcolumn}
\usepackage{bm}
\usepackage{color}
\usepackage{soul}
\usepackage{babel}
\usepackage{amsfonts}
\usepackage{slashed}
\usepackage{enumerate}

\usepackage{babel}

\makeatother

\begin{document}
\global\long\def\sgn{\mathrm{sgn}}%
\global\long\def\ket#1{\left|#1\right\rangle }%
\global\long\def\bra#1{\left\langle #1\right|}%
\global\long\def\sp#1#2{\langle#1|#2\rangle}%
\global\long\def\abs#1{\left|#1\right|}%
\global\long\def\avg#1{\langle#1\rangle}%

\title{Quantum Zeno effect appears in stages}
\author{Kyrylo Snizhko \includegraphics[height=0.3cm]{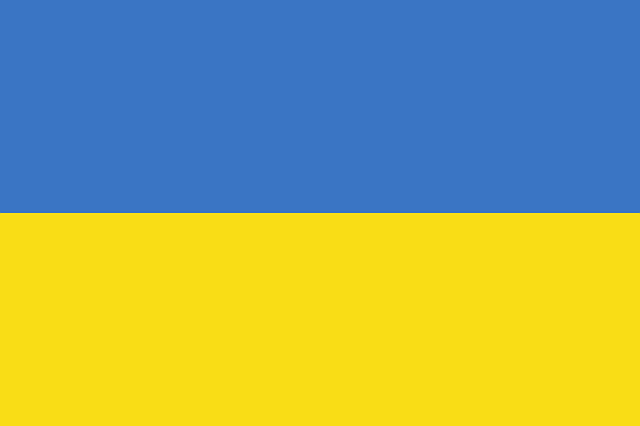}\,\,}
\affiliation{Department of Condensed Matter Physics, Weizmann Institute of Science,
Rehovot, 76100 Israel}
\author{Parveen Kumar \includegraphics[height=0.3cm]{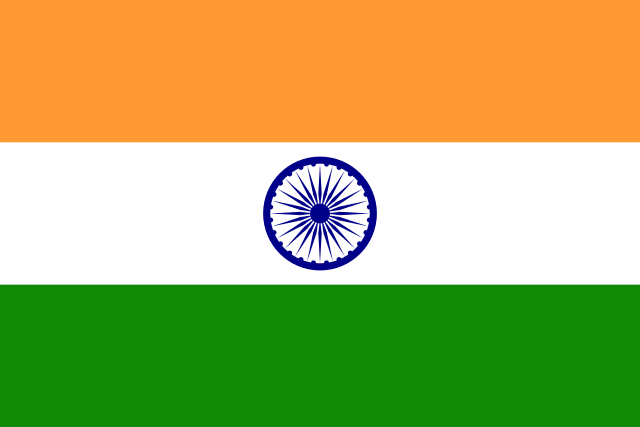}\,\,}
\affiliation{Department of Condensed Matter Physics, Weizmann Institute of Science,
Rehovot, 76100 Israel}
\author{Alessandro Romito \includegraphics[height=0.3cm]{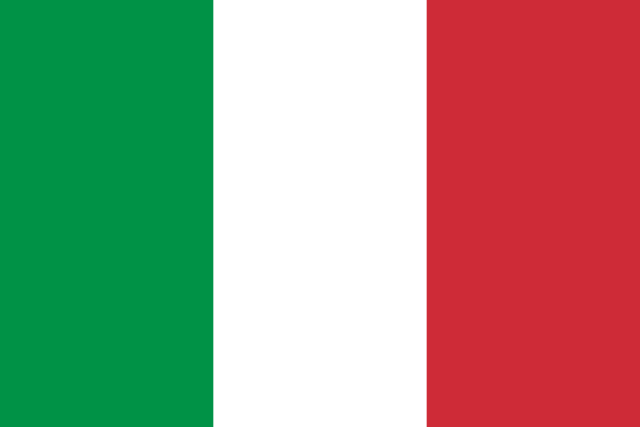}\,\,}
\affiliation{Department of Physics, Lancaster University, Lancaster LA1 4YB, United
Kingdom}
\begin{abstract}
In the quantum Zeno effect, quantum measurements can block the coherent
oscillation of a two level system by freezing its state to one of
the measurement eigenstates. The effect is conventionally controlled
by the measurement frequency. Here we study the development of the
Zeno regime as a function of the measurement strength for a continuous
partial measurement. We show that the onset of the Zeno regime is
marked by a \emph{cascade of transitions} in the system dynamics as
the measurement strength is increased. Some of these transitions are
only apparent in the collective behavior of individual quantum trajectories
and are invisible to the average dynamics. They include the appearance
of a region of dynamically inaccessible states and of singularities
in the steady-state probability distribution of states. These newly
predicted dynamical features, which can be readily observed in current
experiments, show the coexistence of fundamentally unpredictable quantum
jumps with those continuously monitored and reverted in recent experiments.
\end{abstract}
\maketitle
\vspace{0.5pc}

\emph{Introduction}.---The evolution of a quantum system under measurement
is inherently stochastic due to the intrinsic quantum fluctuations
of the detector \citep{Jacobs2014a}. If these fluctuations can be
accurately monitored, measurements can be used to track the stochastic
evolution of the system state, i.e., individual quantum trajectories.
From a theoretical tool to investigate open quantum systems \citep{H.Carmichael1993},
quantum trajectories have become an observable reality in experiments
in optical \citep{Kuhr2007,Dotsenko2011} and solid state \citep{Murch2013,Murch2013a,Weber2014}
systems. Tracking quantum trajectories has been exploited as a tool
to engineering quantum states via continuous feedback control \citep{Vijay2012,Blok2014,DeLange2014}
and entanglement distillation \citep{Riste2013,Roch2014}. It has
been used to observe fundamental properties of quantum measurements
\citep{Groen2013,Campagne-Ibarcq2014,Tan2015,naghiloo2020heat,naghiloo2018information}
and, recently, to predict topological transitions in measurement-induced
geometric phases \citep{Cho2019,Gebhart2020,Snizhko2020b,Snizhko2020c}
and many-body entanglement phase transitions in random unitary circuits,
invisible to the average dynamics \citep{Li2018,Chan2018,Skinner2018,Szyniszewski2019}.
Monitoring quantum trajectories has also made possible anticipating
and correcting quantum jumps in superconducting qubits \citep{Minev2019}.

The above-mentioned transitions stem from the basic physics of the
quantum Zeno effect \citep{Misra1977,Peres1980}. In this regime,
as a result of repeated measurements, the system state is mostly frozen
next to one of the measurement eigenstates, yet rarely performs quantum
jumps between them. The crossover between coherent oscillations and
the Zeno regime is controlled by the frequency of the measurement
and has been extensively explored both theoretically \citep{Facchi2002,Facchi1999a,Burgarth2013,Gherardini2016,Elliott2016a,Majeed2018}
and experimentally \citep{Kwiat1999,Fischer2001,Wolters2013,Signoles2014,Schafer2014}.
Beyond projective measurements, the onset of the Zeno regime is richer
\citep{Layden2015,Zhang2019}, and quantum jumps appear as part of
continuous stochastic dynamics. For example, in a system monitored
via continuous partial measurements, quantum jumps can be anticipated,
continuously monitored, and reverted \citep{Minev2019}, a task which
is fundamentally impossible with projective measurements. Moreover,
the onset of the Zeno regime with non-projective measurements is more
convoluted and has been characterized by different measurement strengths
and phenomenology based on the dynamics of the detector signal \citep{Li2014a},
average \citep{Presilla1996,Gurvitz2003,Koshino2005,Chantasri2013,Kumar2020b},
or postselected \citep{Ruskov2007,Li2020c} state evolution.
\begin{figure}
\begin{centering}
\includegraphics[width=1\columnwidth]{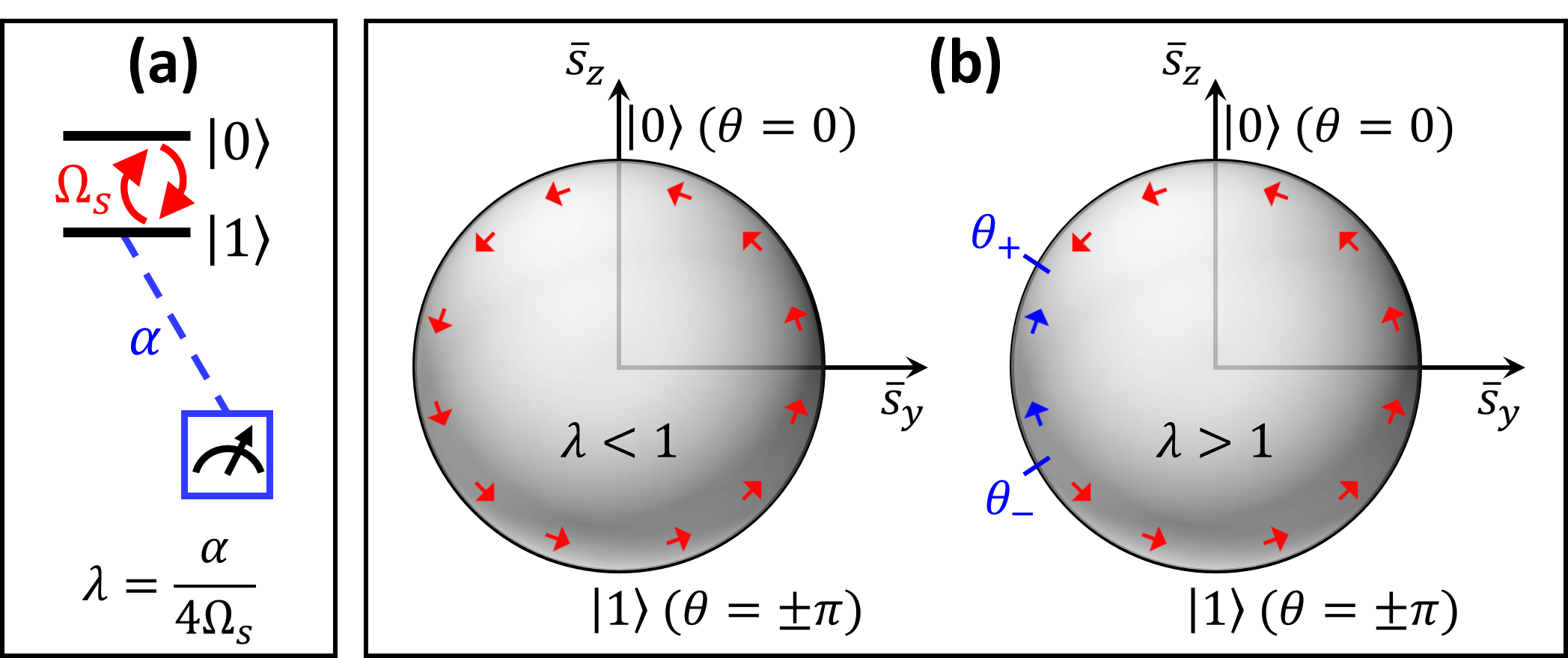}
\par\end{centering}
\caption{\label{fig:circle_flow} (a) The system. A Hamiltonian induces oscillations
between levels $\protect\ket 0$ and $\protect\ket 1$ of a qubit,
which is continuously measured by a detector weakly coupled to one
of the levels. (b) Dynamical flow (red and blue arrows) of $\theta(t)$
from Eq. (\ref{eq:langevin}) under ``no-click'' postselected dynamics.
For sufficiently weak measurements, $\lambda<1$, (left) the dynamics
is oscillatory; for $\lambda>1$ (right), a stable and an unstable
fixed points ($\theta_{+}$ and $\theta_{-}$ respectively) emerge.
The states in the interval $\theta\in(-\pi,\theta_{+})$ are inaccessible
to the system under both the ``no-click'' and the full stochastic
dynamics.}
\end{figure}

Here we study the transition between the regimes of coherent oscillations
and Zeno-like dynamics in a qubit subject to continuous partial measurements,
cf.~Fig.~\ref{fig:circle_flow}(a), a model directly describing
some recent experiments \citep{Minev2019}. By investigating the full
stochastic dynamics of quantum trajectories, we show that the quantum
Zeno regime is established via a \emph{cascade of transitions} in
the system dynamics, some being invisible to the average dynamics.
Furthermore, we find that, in the Zeno regime, catchable continuous
jumps between states $\ket 1$ and $\ket 0$ necessarily have a discontinuous
counterpart, jumps between $\ket 0$ and $\ket 1$, which are inherently
unpredictable in individual realizations. Our results provide a unified
picture of the onset of the Zeno regime arising from continuous partial
measurements and demonstrate that investigating individual quantum
trajectories can uncover drastically new physics even in simple and
well-studied systems. Our findings may be relevant for quantum error
correction protocols employing continuous partial measurements \citep{Chen2020,Kim2012}.

\emph{Model and post-selected dynamics}.---We consider a qubit performing
coherent quantum oscillations between states $\ket 0$ and $\ket 1$
due to the Hamiltonian $H_{s}=\Omega_{s}\sigma_{x}$, where $\Omega_{s}>0$;
at the same time the qubit is monitored by a sequence of measurements
at intervals $dt\ll1/\Omega_{s}$ -- cf.~Fig.~\ref{fig:circle_flow}(a).
Each measurement is characterized by two possible readouts $r=0$
(no-click) and $1$ (click). The corresponding measurement back-action
$\ket{\psi}\to M^{(r)}\ket{\psi}$ is given by the operators
\begin{equation}
M^{(0)}=\ket 0\bra 0+\sqrt{1-p}\ket 1\bra 1,\quad M^{(1)}=\sqrt{p}\ket 1\bra 1,\label{eq:kraus1}
\end{equation}
where $p\in[0,1]$ controls the measurement strength. For $p=1$,
each measurement is projective and this induces the conventional quantum
Zeno effect with the system being frozen in one of the measurement
eigenstates, $\ket 0$ or $\ket 1$. In the opposite limit, $p=0$,
essentially no measurement takes place, and the system performs Rabi
oscillations under $H_{s}$. We investigate the intermediate case
of $p=\alpha dt$ with $dt\to0$, and $\alpha\geq0$ controlling the
effective measurement strength over a finite time interval. A physical
model of this measurement process is realized by coupling the system
to a two-level system detector that, in turn, is subject to projective
measurements, see Appendix~\ref{sec:model} for details.

In each infinitesimal step the measurement and the system evolution
add up to give the combined evolution
\begin{equation}
\ket{\psi(t+dt)}=M^{(r)}U\ket{\psi(t)},\label{eq:evolution-elem}
\end{equation}
where $U=e^{-iH_{s}dt}\approx1-iH_{s}dt$ is the Hamiltonian unitary
evolution over an infinitesimal time interval $dt$. When the system
is initialized in $\ket 0$ or $\ket 1$, its evolution is constrained
to the $y$--$z$ section of the Bloch sphere and the state has the
form $\ket{\psi(t)}=\ket{\psi(\theta(t))}=\cos\frac{\theta(t)}{2}\ket 0+i\sin\frac{\theta(t)}{2}\ket 1$.
Eq.~(\ref{eq:evolution-elem}) translates onto
\begin{equation}
\theta(t+dt)=\begin{cases}
\theta(t)-\Omega(\theta(t))\:dt & {\rm if}\,\,r=0\\
\pi & {\rm if}\,\,r=1
\end{cases},\label{eq:langevin}
\end{equation}
where $\Omega(\theta)=2\Omega_{s}\left[1+\lambda\sin\theta\right]$
and $\lambda=\frac{\alpha}{4\Omega_{s}}$ sets the strength of the
measurement relative to the Hamiltonian. A measurement yielding readout
$r=1$ immediately projects the system onto state $\ket 1$, while
a ``no-click'' $r=0$ readout implies an infinitesimal evolution
of the state with angular velocity $\Omega(\theta)$. The probabilities
of the two possible readouts are given by
\begin{equation}
p_{r=1}\equiv p_{1}=\alpha dt\sin^{2}\frac{\theta(t)}{2},\quad p_{r=0}\equiv p_{0}=1-p_{1}\label{eq:probabilities}
\end{equation}
and depend on the qubit state, i.e. on $\theta(t)$.

For understanding the full stochastic dynamics, it is instructive
to review its continuous ``no-click'' part, previously analyzed
in Ref.~\citep{Ruskov2007}. In this case the state evolution is
governed by the the differential equation $\dot{\theta}=-\Omega(\theta)$.
The corresponding flow of the variable $\theta$ is shown in Fig.~\ref{fig:circle_flow}(b).
Since $\Omega_{s}>0$ and $\lambda>0$, for any $\theta\in(0,\pi)$,
we have $\Omega(\theta)>0$, and the system evolves continuously towards
$\theta=0$. Notably, this is \emph{the only way} for the system state
to evolve from $\ket 1$ to $\ket 0$ and it corresponds to the quantum
jumps that have been continuously monitored in Ref.~\citep{Minev2019}.
The transition from $\ket 0$ to $\ket 1$, instead, takes place via
the region $\theta\in(-\pi,0)$ and has richer dynamics controlled
by the measurement strength. For sufficiently weak measurements, $0\leq\lambda<1$,
one has $\Omega(\theta)>0$ for any $\theta$, and the system monotonously
evolves towards $\theta=-\pi$; however, for $\lambda>1$ there appear
two fixed points, $\Omega(\theta_{\pm})=0$, at
\begin{equation}
\theta_{\pm}=2\arctan\left(-\lambda\pm\sqrt{\lambda^{2}-1}\right),\label{eq:fixed}
\end{equation}
where $\theta_{+}$ is a stable point, while $\theta_{-}$ is an unstable
one, as shown in Fig.~\ref{fig:circle_flow}(b). Under the $r=0$
postselected dynamics for $\lambda>1$, the system will eventually
flow to $\theta=\theta_{+}$ \citep{Ruskov2007} (where it remains
until the occurrence of a click, which collapses the system to $\ket 1$).

\begin{figure*}
\begin{centering}
\includegraphics[width=1\textwidth]{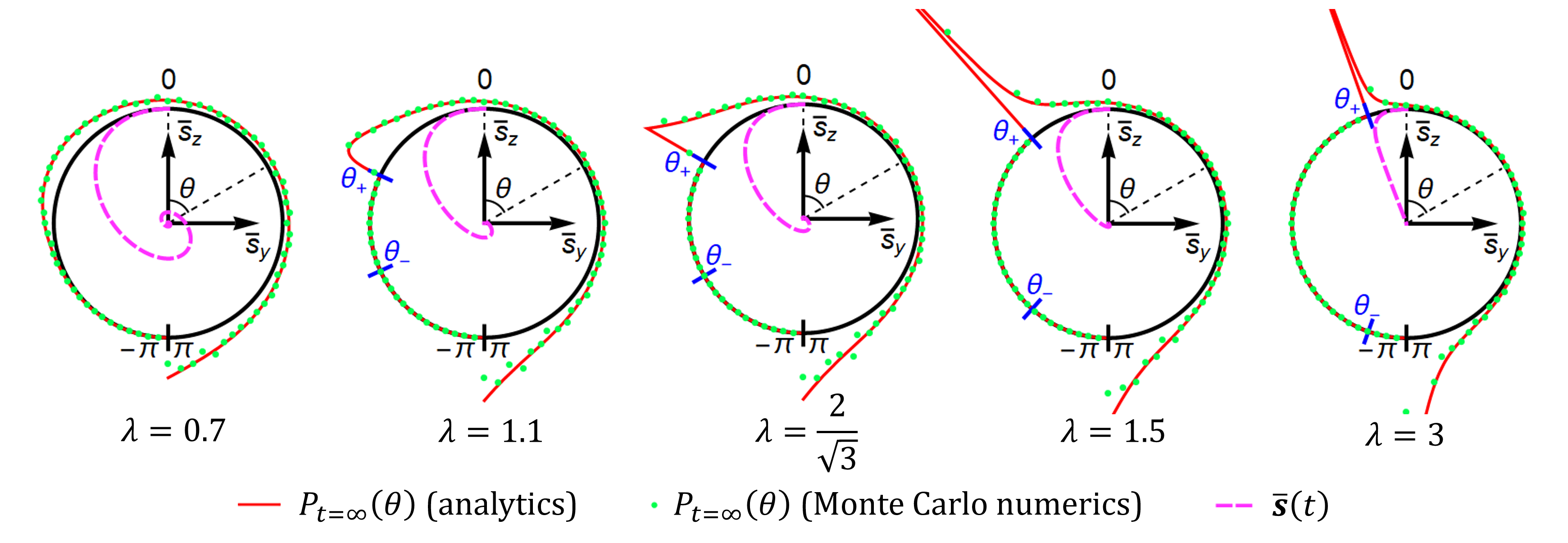}
\par\end{centering}
\caption{\label{fig:detection+Ptheta} The stochastic dynamics of the qubit
state in the $y$--$z$ section of the Bloch sphere exhibits transitions
at relative measurement strengths $\lambda=1$, $2/\sqrt{3}$, and
$2$. The long-time probability distribution $P_{t=\infty}(\theta)$
at different values of $\lambda$ is shown as the height above the
unit circle for the analytic result (red solid line) and for the numerical
simulations (green dots). The trajectory of the expectation values
$\bar{s}_{y,z}(t)$, for the system initialized in $\protect\ket 0$
at $t=0$, are shown with the dashed magenta lines. Each numerical
simulation involved $10000$ stochastic trajectory realizations, tracing
the evolution until $t=10$, with measurements yielding random outcomes
happening at intervals $dt=0.01$, using $\Omega_{s}=1$; $\theta$
is binned in intervals of size $5{^\circ}$.}
\end{figure*}

\emph{Stochastic evolution and dynamical transitions}.---Beyond the
postselected $r=0$ quantum trajectory, the stochastic dynamics of
the system is described by the probability density $P_{t}(\theta)$
of being in the state $\ket{\psi(\theta)}$ at time $t$. Using Eqs.~(\ref{eq:langevin}--\ref{eq:probabilities}),
one derives the master equation for $P_{t}(\theta)$
\begin{multline}
\frac{dP_{t}(\theta)}{dt}=\biggl[\partial_{\theta}\left(\Omega(\theta)P_{t}(\theta)\right)\\
-4\Omega_{s}\lambda\sin^{2}\frac{\theta}{2}P_{t}(\theta)+4\Omega_{s}\lambda\delta(\theta-\pi)\int_{0}^{2\pi}d\tilde{\theta}\sin^{2}\frac{\tilde{\theta}}{2}P_{t}(\tilde{\theta})\biggr].\label{eq:-diff-_equation_no_noise}
\end{multline}
Here, the first term on the r.h.s. describes the ``no-click'' evolution,
the second term describes the reduction of $P_{t}(\theta)$ due to
clicks that happen with probability $p_{1}=4\Omega_{s}\lambda\sin^{2}\frac{\theta}{2}dt$,
cf.~Eq.~(\ref{eq:probabilities}), while the last term accounts
for the clicks bringing the states from any $\theta$ to $\theta=\pi$.

Two experimentally accessible quantities directly related to $P_{t}(\theta)$
capture the main physics: the steady-state distribution $P_{\infty}(\theta)\equiv\lim_{t\to\infty}P_{t}(\theta)$,
and the average ``polarization'' of the qubit, $\bar{\boldsymbol{s}}(t)\equiv(\bar{s}_{y}(t),\bar{s}_{z}(t))$,
where $\bar{s}_{i}(t)\equiv\langle\sigma_{i}(t)\rangle=\int_{-\pi}^{\pi}\bra{\psi(\theta)}\sigma_{i}\ket{\psi(\theta)}P_{t}(\theta)\,d\theta$,
$i=y,z$. Both quantities are plotted in Fig.~\ref{fig:detection+Ptheta}.
They showcase three qualitative transitions in the dynamics as function
of the measurement strength.

We can readily present the \emph{key physics of these transitions}
before entering all the features in due details. For sufficiently
small $\lambda,$ the qubit can be found in any state with finite
probability density $P_{\infty}(\theta)\neq0$. In particular, it
is possible to evolve from $\ket 0$ to $\ket 1$ via trajectories
involving a detector click as well as no clicks from the detector.
Instead, the evolution from $\ket 1$ to $\ket 0$ happens only via
no click sequences, as noted above. The first and most drastic transition
happens at $\lambda=1$, above which there opens a region of $\theta\in(-\pi;\theta_{+}]$
where $P_{\infty}(\theta)=0$. In fact, this region is inaccessible
for the qubit at any time $t$, hence, for $\lambda>1$ all quantum
trajectories from $\ket 0$ to $\ket 1$ must involve a detector click.
Generically, the click may occur when the qubit has not reached $\theta_{+}$,
which is typically the case. The second transition happens at $\lambda=2/\sqrt{3}$,
above which $P_{\infty}(\theta)$ diverges at $\theta=\theta_{+}$.
This indicates that the system initialized in $\ket 0$ typically
reaches the vicinity of $\theta_{+}$, and spends a long time there,
before the click and the corresponding jump to $\theta=\pi$ take
place. So far, the population imbalance between $\ket 0$ and $\ket 1$,
$\bar{s}_{z}(t)$, exhibits oscillations, which are reflected in the
oscillations of the average state polarization, $\bar{\boldsymbol{s}}(t)$.
The third and final transition at $\lambda=2$ marks the end of the
oscillations, so that, for $\lambda>2$, $\bar{s}_{z}(t)$ steadily
decays in time, completing the final onset of Zeno-like dynamics.
These transitions set the overall picture of the onset of Zeno regime
in the system, and constitute the main findings of our work.

To analyze these transitions and their implications in some detail,
consider first the non-trivial steady state, $P_{\infty}(\theta)$.
From the condition $dP_{t\rightarrow\infty}(\theta)/dt=0$, Eq.~(\ref{eq:-diff-_equation_no_noise})
gives
\begin{equation}
P_{{\rm \infty}}(\theta)=\frac{\lambda\exp\left[\frac{2\lambda}{\sqrt{1-\lambda^{2}}}\left(\arctan\frac{\lambda+\tan\frac{\theta}{2}}{\sqrt{1-\lambda^{2}}}-\frac{\pi}{2}\right)\right]}{(1+\lambda\sin\theta)^{2}\left[1-\exp\left(-\frac{2\pi\lambda}{\sqrt{1-\lambda^{2}}}\right)\right]},\label{eq:steady}
\end{equation}
for $\lambda<1$, while for $\lambda>1$ the expression reads
\begin{equation}
P_{{\rm \infty}}(\theta)=\begin{cases}
\frac{\lambda}{(1+\lambda\sin\theta)^{2}}\left(\frac{\tan\frac{\theta}{2}+\lambda-\sqrt{\lambda^{2}-1}}{\tan\frac{\theta}{2}+\lambda+\sqrt{\lambda^{2}-1}}\right)^{\frac{\lambda}{\sqrt{\lambda^{2}-1}}}, & \theta\in(\theta_{+};\pi],\\
0, & \theta\in(-\pi;\theta_{+}].
\end{cases}\label{eq:steady-strong}
\end{equation}
In Fig.~\ref{fig:detection+Ptheta}, the analytical results in Eqs.
(\ref{eq:steady}, \ref{eq:steady-strong}) are compared with Monte
Carlo numerical simulations of individual quantum trajectories, showing
excellent agreement. The first two transitions in the system dynamics
highlighted above, are evident from $P_{\infty}(\theta)$ -- cf.~Fig.~\ref{fig:detection+Ptheta}.
The opening of the forbidden region $(-\pi;\theta_{+}]$ appears discontinuously
rather than opening smoothly, since $\theta_{+}=\theta_{-}=-\pi/2$
at $\lambda=1$. It manifests itself in the non-analytic behavior
of $P_{{\rm \infty}}(\theta)$ as a function of $\lambda$ at $\lambda=1$.
For $\lambda>1$, the second transition shows up at $\theta\approx\theta_{+}$,
where
\begin{equation}
P_{{\rm \infty}}(\theta)\propto\left(\tan\frac{\theta}{2}-\tan\frac{\theta_{+}}{2}\right)^{\frac{\lambda}{\sqrt{\lambda^{2}-1}}-2},
\end{equation}
which diverges at $\theta=\theta_{+}$ for $\lambda>2/\sqrt{3}$.
The physics underpinning this transition is evinced by Eq.~(\ref{eq:-diff-_equation_no_noise})
at $\theta=\theta_{+}$. The first two terms in the r.h.s., $P_{t}(\theta_{+})\partial_{\theta_{+}}\Omega(\theta_{+})$
and $-4\Omega_{s}\lambda\sin^{2}\frac{\theta_{+}}{2}P_{t}(\theta_{+})$,
describe the rate of accumulation of probability for states at $\theta\approx\theta_{+}$
due to no-click dynamics and the loss of such probability due to detector
clicks respectively. At $\lambda=2/\sqrt{3}$ the two terms balance
each other, and for $\lambda>2/\sqrt{3}$ the former dominates. Note
that the $\lambda=2/\sqrt{3}$ transition goes unnoticed in both the
average (state polarization $\bar{\boldsymbol{s}}(t)$) behavior and
in the post-selected $r=0$ dynamics.

\begin{figure}
\begin{centering}
\includegraphics[width=1\columnwidth]{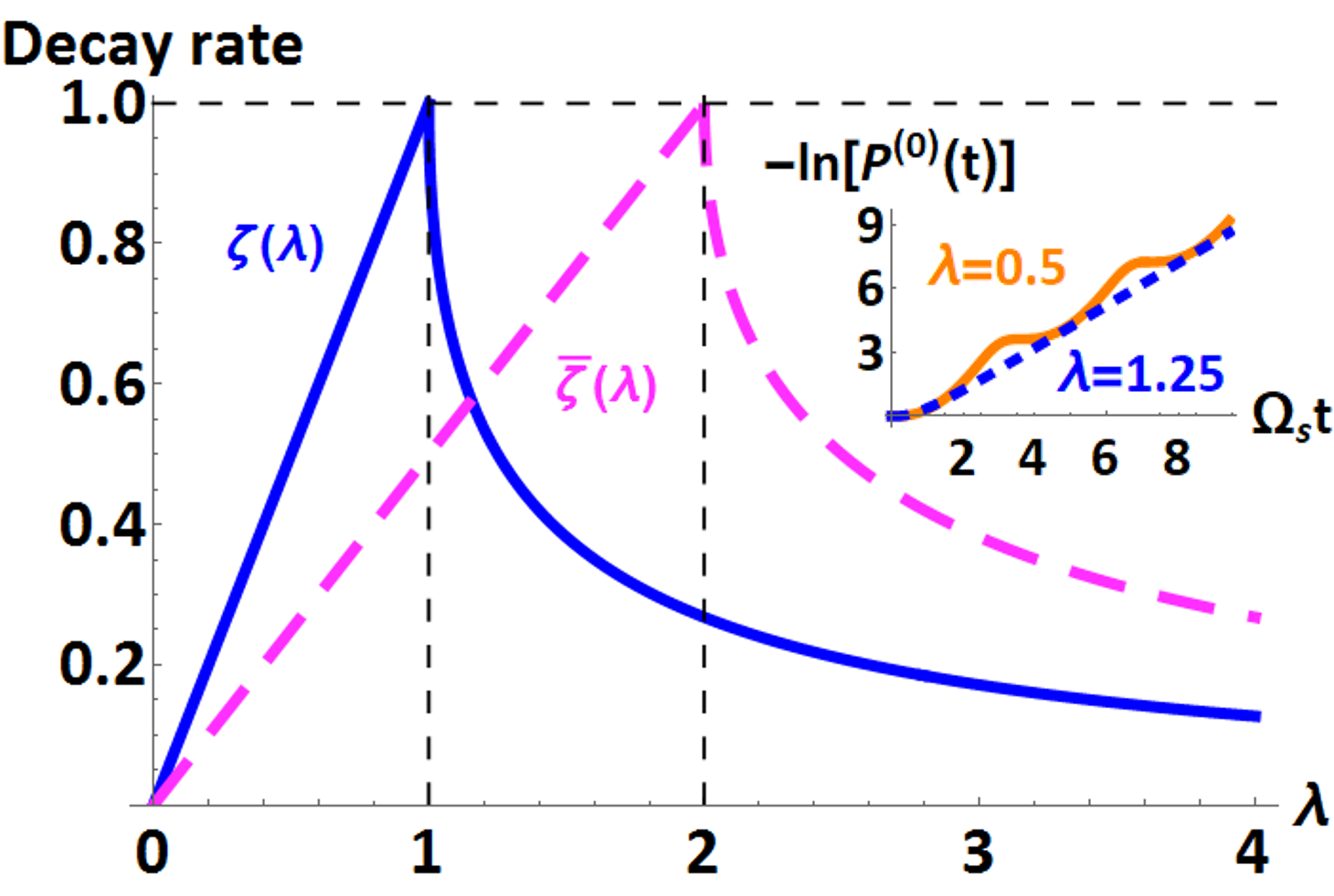}
\par\end{centering}
\caption{\label{fig:P0_decay_properties}The decay rates $\zeta(\lambda)$
(solid) and $\bar{\zeta}(\lambda)$ (dashed) characterizing respectively
the probability to observe no clicks $P^{(0)}(t)$ and the survival
probability $(1+\bar{s}_{z}(t))/2$. Note the respective decay rate
maxima at $\lambda=1$ and $\lambda=2$. Inset---The time dependence
of $P^{(0)}(t)$ for $\lambda=0.5$ (solid) and $\lambda=1.25$ (dashed).
At long times $P^{(0)}(t)$ decays exponentially with (without) superimposed
oscillations for $\lambda<1$ ($\lambda>1$).}
\end{figure}

While the steady-state properties of $P_{\infty}(\theta)$ showcase
these transitions, their \emph{role} in the onset of the Zeno regime
is fully unveiled only in the full stochastic \emph{dynamics}. To
appreciate that, consider the probability $P^{(0)}(t)$ of obtaining
a sequence of no clicks ($r=0$ readouts) of duration $t$, cf.~Fig.~\ref{fig:P0_decay_properties}
(inset). We obtain that at long-times $P^{(0)}(t)$ decays exponentially
as
\begin{equation}
P^{(0)}(t)\propto e^{-2\Omega_{s}\zeta(\lambda)t}\left[A(\lambda)+B(\lambda)\cos(\varepsilon t+\varphi(\lambda))\right],\label{eq:no-click_prob_asymptotics}
\end{equation}
where $B(\lambda)=0$ for $\lambda>1$, the frequency of the oscillatory
term for $\lambda<1$ is $\varepsilon=2\Omega_{s}\sqrt{1-\lambda^{2}}$,
and the decay rate is $\zeta(\lambda)={\rm Re}\left[\lambda-\sqrt{\lambda^{2}-1}\right]$;
see Appendix~\ref{sec:postselected} for the derivation. The value
$\lambda=1$ is special in two respects. For $\lambda<1$, the system
state rotates between $\ket 0$ and $\ket 1$ under no-click dynamics.
However, the probability of observing a click is different for different
$\theta$ --- thence the oscillations of $P^{(0)}(t)$. With the
appearance of the forbidden region, the evolution under no clicks
readout is frozen at $\theta=\theta_{+}$, hence the oscillations
of $P^{(0)}(t)$ disappear. The less obvious effect is that at $\lambda=1$
the decay rate $\zeta(\lambda)$ is maximal, cf.~Fig.~\ref{fig:P0_decay_properties}.
Therefore, the probability of observing a long sequence of ``no-clicks''
increases with the measurement strength for $\lambda\geq1$, while
it decreases for $\lambda\leq1$.

Consider now the probability $P^{(0)}(\theta)$ to reach a particular
value of $\theta$ under no-click dynamics. $P^{(0)}(\theta)$ is
obtained from $P^{(0)}(t)$ and the no-click evolution evolution $\dot{\theta}(t)=-\Omega(\theta(t))$
via $P^{(0)}(\theta_{0})\equiv P^{(0)}(t_{0})$ with $t_{0}$ satisfying
$\theta(t_{0})=\theta_{0}$. In proximity of $\theta=\theta_{+}$,
one has, cf.~Appendix~\ref{sec:postselected},
\begin{equation}
P^{(0)}(\theta\approx\theta_{+})\propto\left(\tan\frac{\theta}{2}-\tan\frac{\theta_{+}}{2}\right)^{\frac{\lambda}{\sqrt{\lambda^{2}-1}}-1}.
\end{equation}
Note that $P^{(0)}(\theta)$ vanishes at $\theta=\theta_{+}$ for
any finite $\lambda>1$. However, $dP^{(0)}(\theta)/d\theta$ vanishes
at $\theta_{+}$ for $\lambda<2/\sqrt{3}$ and diverges for $\lambda>2/\sqrt{3}$.
Therefore, for $\lambda<2/\sqrt{3}$, the system typically jumps to
$\theta=\pi$ via a detector click before it reaches $\theta_{+}$.
For $\lambda>2/\sqrt{3}$, the system is likely to reach a close vicinity
of $\theta_{+}$ before a click happens.

To observe the last transition, $\lambda=2$, one needs to consider
the average state polarization, $\bar{\boldsymbol{s}}(t)=(\bar{s}_{y}(t),\bar{s}_{z}(t))$,
and in particular the population imbalance, $\bar{s}_{z}(t)$. When
the system is initialized in state $\ket 0$ at time $t=0$, using
Eq.~(\ref{eq:-diff-_equation_no_noise}), one finds, cf.~Appendix~\ref{sec:postselected},
\begin{equation}
\bar{s}_{z}(t)=e^{-\Omega_{s}\lambda t}\left(\cosh\Omega_{s}t\sqrt{\lambda^{2}-4}+\lambda\frac{\sinh\Omega_{s}t\sqrt{\lambda^{2}-4}}{\sqrt{\lambda^{2}-4}}\right).\label{eq:density-matrix}
\end{equation}
One sees that $\lambda=2$ marks a transition from oscillatory (at
$\lambda<2$) to non-oscillatory (at $\lambda>2$) dynamics. The same
transition is observed in $\bar{s}_{y}(t)$. Similarly to $P^{(0)}(t)$
in Eq.~(\ref{eq:no-click_prob_asymptotics}), $\bar{s}_{z}(t\rightarrow+\infty)\propto e^{-2\Omega_{s}\bar{\zeta}(\lambda)t}$
with the decay rate $\bar{\zeta}(\lambda)=\mathrm{Re}\left[\lambda-\sqrt{\lambda^{2}-4}\right]/2$.
Therefore, the decay rate exhibits a maximum at $\lambda=2$, cf.~Fig.~\ref{fig:P0_decay_properties}.
Importantly, $\bar{s}_{z}(t)$ characterizes not only the population
imbalance but also the survival probability $(1+\bar{s}_{z}(t))/2$,
i.e., the probability to find the system in state $\ket 0$ when performing
a projective measurement at time $t$. The decay rate behavior implies
that the long-time survival probability increases with increasing
$\lambda$ when $\lambda\geq2$, and decreases otherwise \footnote{In fact, it follows from Eq.~(\ref{eq:density-matrix}) that $\partial_{\lambda}\bar{s}_{z}(t)\geq0$
for any $t\geq0$ when $\lambda>2$ .}. On this ground, one marks the transition at $\lambda=2$ as the
final onset of the Zeno-like dynamics. Notably, the $\lambda=2$ transition
is also reflected in a topological transition in the statistics of
the detector clicks~\citep{Li2014a}.

We would like to note that while we used different quantities to showcase
each of the three dynamical transitions, all three of them can be
inferred by looking at a single quantity: the eigenmode spectrum of
Eq.~(\ref{eq:-diff-_equation_no_noise}). This more mathematical
identification of the transitions and its relation to the physics
described here is discussed in Appendix~\ref{sec:master}.

\emph{Observing the transitions experimentally}.---The above physics
can be readily observed in the setup of recent experiments \citep{Minev2019}
by adjusting the measurement strength/Rabi frequency. The simplest
transition to observe is that at $\lambda=2$, which is apparent in
routine experiments measuring the average polarization or survival
probability in $\ket 0$. Observing the transitions at $\lambda=1$
and $\lambda=2/\sqrt{3}$ requires sampling the distribution $P_{\infty}(\theta)$
by tracking individual quantum trajectories for sufficiently long
times and performing quantum tomography on the final states. This
is possible with state-of-the-art experimental techniques \citep{Murch2013,Minev2019},
though laborious. A somewhat less laborious alternative to observe
the $\lambda=1$ transition is to measure the probability to observe
no clicks for a given time, $P^{(0)}(t).$ While this requires recording
every measurement outcome, it does not require knowing the qubit state
at time $t$. For the transtion at $\lambda=2/\sqrt{3}$, one can
measure the probability to reach a specific $\theta$ by a sequence
of no clicks, $P^{(0)}(\theta)$, which requires further tracing the
qubit state up to time $t$. This can be done either by inferring
the state from the theoretical dependence $\theta(t)$ or via a tomography
of states \emph{postselected} on $r=0$ readouts (which has been implemented
in Ref.~\citep{Minev2019} for $\lambda\gg1$).

\emph{Conclusions}.---Here we have studied the full stochastic dynamics
of a system subject to a constant Hamiltonian and a continuous partial
measurement. We have shown that the onset of Zeno-like regime is preceded
by a number of drastic qualitative changes in the system dynamics.
Each such transition introduces a different feature of the fully localized
dynamics, starting with the opening up of a finite size region of
forbidden states, followed by a singularity in the steady-state probability
distribution of states, and ultimately a non-oscillatory dynamics
of the qubit survival probability. We have proposed how to observe
our findings in current experiments. Strikingly, depending on the
definition of ``Zeno-like regime'', one could call each of the transitions
its onset. For example, the probability of observing a long sequence
of ``no clicks'' starts increasing with increasing the measurement
strength at $\lambda=1$. The survival probability starts increasing
with increasing the measurement strength only after the last transition
at $\lambda=2$. Some of our findings may depend on the specific measurement
model, making it of interest to study the onset of the Zeno regime
beyond continuous partial measurement.
\begin{acknowledgments}
We thank Serge Rosenblum and Fabien Lafont for their comments on the
manuscript. K.\,S. and P.\,K. have contributed equally to this work.
K.\,S. and P.\,K. acknowledge funding by the Deutsche Forschungsgemeinschaft
(DFG, German Research Foundation) -- Projektnummer 277101999 --
TRR 183 (project C01) and Projektnummer EG 96/13-1, as well as by
the Israel Science Foundation (ISF). A.\,R. acknowledges EPSRC via
Grant No. EP/P010180/1.
\end{acknowledgments}

\bibliography{zeno-dark-bib}

\appendix
\onecolumngrid

\section{\label{sec:model}A physical model of the measurement}

The system under consideration in the manuscript is a qubit ($\ket 0$,
$\ket 1$) evolving under its own Hamiltonian and being measured by
a two-state detector ($\ket{0_{d}}$, $\ket{1_{d}}$) at intervals
$dt$. The system's Hamiltonian is
\begin{equation}
H_{s}=\Omega_{s}\sigma_{x}^{(s)}.
\end{equation}
We consider a system-detector Hamiltonian given by
\begin{equation}
H_{s-d}=\frac{J}{2}(1-\sigma_{z}^{(s)})\sigma_{y}^{(d)},
\end{equation}
where the detector is also assumed to be a two-level system. The detector
is initially prepared in the state $\ket{0_{d}}$ for each measurement,
i.e. at the beginning of each time step. The system's evolution under
the combined effect of its Hamiltonian and the coupling to the detector
is given by the unitary evolution due to
\begin{equation}
H=H_{s}+H_{s-d},
\end{equation}
for time $dt$, after which the detector is read out with readouts
$r=0,1$ corresponding to it being in $\ket{r_{d}}$.

We consider the scaling limit of continuous measurements defined as
$dt\rightarrow0$, $J^{2}dt\rightarrow\alpha=\mathrm{const}$ (i.e.,
$J=\sqrt{\alpha/dt}$). In this limit, the measurement and the system
evolution do not intermix in a single step, therefore,
\begin{equation}
\ket{\psi(t+dt)}=M^{(r)}U_{s}\ket{\psi(t)}.
\end{equation}
The unitary evolution due to the system's Hamiltonian is
\[
U_{s}=e^{-iH_{s}dt}=\cos\Omega_{s}dt-i\sigma_{x}^{(s)}\sin\Omega_{s}dt=\begin{pmatrix}1 & -i\Omega_{s}dt\\
-i\Omega_{s}dt & 1
\end{pmatrix}+O(dt^{2}).
\]
The measurement back action matrices, combining the effects of the
system-detector evolution and the readout in the state $\ket{r_{d}}$,
defined as
\begin{equation}
M^{(r)}=\bra{r_{d}}e^{-iH_{s-d}dt}\ket{0_{d}},
\end{equation}
are
\begin{equation}
M^{(0)}=\begin{pmatrix}1 & 0\\
0 & \cos Jdt
\end{pmatrix}=\begin{pmatrix}1 & 0\\
0 & 1-\frac{1}{2}\alpha dt
\end{pmatrix}+O(dt^{2}),
\end{equation}
\begin{equation}
M^{(1)}=\begin{pmatrix}0 & 0\\
0 & \sin Jdt
\end{pmatrix}=\begin{pmatrix}0 & 0\\
0 & \sqrt{\alpha dt}
\end{pmatrix}+O(dt^{3/2}).
\end{equation}
These are the operators in Eq. (\ref{eq:kraus1}) of the manuscript
describing the effect of the measurements.

\section{\label{sec:postselected}Calculation of postselection and survival
probabilities}

In the main text we have presented some of the observable signatures
of the transition in terms of the probabilities $P^{(0)}(t)$ to observe
a no-click sequence of duration $t$ and $P^{(0)}(\theta)$ to reach
$\theta$ by a sequence of no clicks, as well as the behavior of the
average state polarization $\bar{\bm{s}}(t)$. Here we derive the
results stated in the main text.

\subsection{Probabilities $P^{(0)}(t)$ and $P^{(0)}(\theta)$}

To determine the probability of the $r=0$-postselected trajectory,
we start by solving the state evolution under a sequence of $0$ readouts.
The corresponding equation for $\theta$ (cf. Eq. (3) in the manuscript)
is
\begin{equation}
\frac{d\theta}{dt}=-2\Omega_{s}(1+\lambda\sin\theta).
\end{equation}
The solution is:
\begin{equation}
\tan\frac{\theta}{2}=\sqrt{\lambda^{2}-1}\tanh\left(\Omega_{s}\sqrt{\lambda^{2}-1}(t-t_{0})\right)-\lambda.
\end{equation}
Setting the initial condition $\theta(t=0)=0$ and simplifying the
expression, we arrive to
\begin{equation}
\tan\frac{\theta(t)}{2}=\sqrt{\lambda^{2}-1}\coth\left(\Omega_{s}t\sqrt{\lambda^{2}-1}-\frac{1}{2}\ln\frac{\tan\frac{\theta_{+}}{2}}{\tan\frac{\theta_{-}}{2}}\right)-\lambda=-\frac{1}{\lambda+\sqrt{\lambda^{2}-1}\coth\left(\Omega_{s}t\sqrt{\lambda^{2}-1}\right)}.\label{eq:soluzione}
\end{equation}
For $\lambda>1$, this expression describes the evolution of $\theta(t)$
from $\pi$ at $t=-\frac{1}{\Omega_{s}\sqrt{\lambda^{2}-1}}\mathrm{arccoth}\left(\frac{\lambda}{\sqrt{\lambda^{2}-1}}\right)$
to $\theta_{+}$ at $t=+\infty$. For $\lambda<1$, Eq. (\ref{eq:soluzione})
becomes
\begin{equation}
\tan\frac{\theta(t)}{2}=-\frac{1}{\lambda+\sqrt{1-\lambda^{2}}\cot\left(\Omega_{s}t\sqrt{1-\lambda^{2}}\right)}
\end{equation}
and describes the periodic evolution of $\theta(t)$ with period $T=\frac{\pi}{\Omega_{s}\sqrt{\lambda^{2}-1}}$.

We are interested in the the probability of having zero readout at
time $t$, $P_{0}(t)$. Knowing the probability of obtaining $r=0$
in each infinitesimal step, the equation for $P_{0}(t)$ is readily
determined:
\begin{multline}
\frac{dP^{(0)}(t)}{dt}=-\frac{p_{1}}{dt}P^{(0)}(t)=-\alpha\sin^{2}\frac{\theta(t)}{2}P^{(0)}(t)\\
=-\alpha\frac{\tan^{2}\frac{\theta(t)}{2}}{1+\tan^{2}\frac{\theta(t)}{2}}P^{(0)}(t)=-\alpha\frac{\sinh^{2}(\Omega_{s}t\sqrt{\lambda^{2}-1})}{\lambda^{2}\cosh(2\Omega_{s}t\sqrt{\lambda^{2}-1})-1+\lambda\sqrt{\lambda^{2}-1}\sinh(2\Omega_{s}t\sqrt{\lambda^{2}-1})}.
\end{multline}
Integrating the equation and demanding $P_{0}(t=0)=1$, one obtains,
\begin{equation}
P^{(0)}(t)=e^{-2\Omega_{s}\lambda t}\frac{\lambda^{2}\cosh(2\Omega_{s}t\sqrt{\lambda^{2}-1})-1+\lambda\sqrt{\lambda^{2}-1}\sinh(2\Omega_{s}t\sqrt{\lambda^{2}-1})}{\lambda^{2}-1}.
\end{equation}
This expression is \emph{not} singular at $\lambda=1$, where it becomes
$P_{0}(t)=(1+2\Omega_{s}t+2\Omega_{s}^{2}t^{2})e^{-2\Omega_{s}t}$.
For $\lambda<1$, it reduces to
\begin{equation}
P^{(0)}(t)=e^{-2\Omega_{s}\lambda t}\frac{\lambda^{2}\cos(2\Omega_{s}t\sqrt{1-\lambda^{2}})-1-\lambda\sqrt{1-\lambda^{2}}\sin(2\Omega_{s}t\sqrt{1-\lambda^{2}})}{\lambda^{2}-1}.\label{eq:p0}
\end{equation}
From Eq. (\ref{eq:p0}), one can directly derive the long-time behavior
of $P^{(0)}(t)$ reported in the manuscript {[}cf. Eq. (10) therein{]},
which is
\begin{equation}
P^{(0)}(t\rightarrow+\infty)\propto\exp\left(-2\Omega_{s}\lambda t\right)\times\text{oscillating function},
\end{equation}
for $\lambda<1$, and
\begin{equation}
P^{(0)}(t\rightarrow+\infty)\propto\exp\left(-2\Omega_{s}\left[\lambda-\sqrt{\lambda^{2}-1}\right]t\right)
\end{equation}
for $\lambda>1$.

The probability to reach state $\theta$ via a sequence of $r=0$
readouts, $P^{(0)}(\theta)$ is obtained from the equation
\begin{equation}
\frac{dP_{0}(\theta)}{d\theta}=\frac{dP_{0}(\theta(t))}{dt}/\frac{d\theta}{dt}=\frac{2\lambda\sin^{2}\frac{\theta}{2}}{1+\lambda\sin\theta}P_{0}(\theta).
\end{equation}
The solution with $P^{(0)}(\theta=0)=1$ is
\begin{multline}
P^{(0)}(\theta)=\frac{1}{1+\lambda\sin\theta}\left(\frac{\tan\frac{\theta}{2}-\tan\frac{\theta_{+}}{2}}{\tan\frac{\theta}{2}-\tan\frac{\theta_{-}}{2}}\right)^{\frac{\lambda}{\sqrt{\lambda^{2}-1}}}\left(\frac{\tan\frac{\theta_{+}}{2}}{\tan\frac{\theta_{-}}{2}}\right)^{-\frac{\lambda}{\sqrt{\lambda^{2}-1}}}\\
=\frac{1}{1+\lambda\sin\theta}\exp\left(\frac{2\lambda}{\sqrt{1-\lambda^{2}}}\left[\arctan\frac{\lambda+\tan\frac{\theta}{2}}{\sqrt{1-\lambda^{2}}}-\arctan\frac{\lambda}{\sqrt{1-\lambda^{2}}}\right]\right).
\end{multline}
When $\lambda>1$, starting at $\theta=0$ it is possible to reach
only the states with $\theta\in(\theta_{+};0]$, which is reflected
in the vanishing of
\begin{equation}
P^{(0)}(\theta\approx\theta_{+})\propto\left(\tan\frac{\theta}{2}-\tan\frac{\theta_{+}}{2}\right)^{\frac{\lambda}{\sqrt{\lambda^{2}-1}}-1},
\end{equation}
and in the properties of its derivatives discussed in the manuscript.

\subsection{\label{subsec:average_polarization}Survival probability and the
average state polarization}

The final quantity used in the manuscript to describe the dynamics
of the system is the average state polarization after time $t$, $\bar{\boldsymbol{s}}(t)\equiv(\bar{s}_{y}(t),\bar{s}_{z}(t))$,
which is related to the survival probability in the initial state
$\ket 0$, $\mathcal{P}(t)=(1+\bar{s}_{z}(t))/2$, i.e. the probability
to measure the system in $\ket 0$ upon a projective measurement at
time $t$. These quantities require averaging over all the state trajectories,
and no postselection is required. The average $\bar{s}_{y}$ and $\bar{s}_{z}$
components of the polarization can be expressed through the state
distribution $P_{t}(\theta)$:
\begin{align}
\bar{s}_{y}(t) & =\int_{-\pi}^{\pi}d\theta P_{t}(\theta)\sin\theta,\\
\bar{s}_{z}(t) & =\int_{-\pi}^{\pi}d\theta P_{t}(\theta)\cos\theta.
\end{align}
It then follows from Eq.~(\ref{eq:-diff-_equation_no_noise}) that
\begin{equation}
\frac{d}{dt}\begin{pmatrix}\bar{s}_{y}\\
\bar{s}_{z}
\end{pmatrix}=\begin{pmatrix}-\frac{\alpha}{2} & -2\Omega_{s}\\
2\Omega_{s} & 0
\end{pmatrix}\begin{pmatrix}\bar{s}_{y}\\
\bar{s}_{z}
\end{pmatrix}.
\end{equation}
The evolution has two eigenvalues, $\Omega_{s}(-\lambda\pm\sqrt{\lambda^{2}-4})$.
When the system is initialized in state $\ket 0$ at $t=0$, the evolution
of $\bar{\boldsymbol{s}}(t)$ is given by
\begin{align}
\bar{s}_{y}(t) & =-2e^{-\Omega_{s}\lambda t}\frac{\sinh\Omega_{s}t\sqrt{\lambda^{2}-4}}{\sqrt{\lambda^{2}-4}},\\
\bar{s}_{z}(t) & =e^{-\Omega_{s}\lambda t}\left(\cosh\Omega_{s}t\sqrt{\lambda^{2}-4}+\frac{\lambda}{\sqrt{\lambda^{2}-4}}\sinh\Omega_{s}t\sqrt{\lambda^{2}-4}\right).
\end{align}
Its long-time behaviour is given by
\begin{equation}
\bar{s}_{y,z}(t)\propto\begin{cases}
\exp\left(-\Omega_{s}\lambda t\right)\times\text{oscillating function}, & \text{for }\lambda<2,\\
\exp\left(-\Omega_{s}\left[\lambda-\sqrt{\lambda^{2}-4}\right]t\right), & \text{for }\lambda>2.
\end{cases}
\end{equation}
 Therefore, at long times, the survival probability $\mathcal{P}(t)$
decays to the steady state value $\mathcal{P}(t\rightarrow\infty)=1/2$.
The decay rate is, $\Omega_{s}\left(\lambda-\sqrt{\lambda^{2}-4}\right)=2\Omega_{s}\left(\frac{\lambda-\sqrt{\lambda^{2}-4}}{2}\right)$.
It exhibits a maximum at $\lambda=2$, resembling the maximum of the
decay rate of $P_{0}(t)$ at $\lambda=1$.

Note that the equation for the polarization evolution, yielding two
eigenvalues, $\Omega_{s}\left(\lambda\pm\sqrt{\lambda^{2}-4}\right)$,
has been obtained from Eq.~(\ref{eq:-diff-_equation_no_noise}).
Thus, Eq.~(\ref{eq:-diff-_equation_no_noise}) ``knows'' about
the eigenvalues $\gamma=\frac{1}{2}\left(-\lambda\pm\sqrt{\lambda^{2}-4}\right)$
for any values of $\lambda$. At the same time, these eigenvalues
correspond to normalizable eigenmodes of Eq.~(\ref{eq:-diff-_equation_no_noise})
only when $\lambda<2/\sqrt{3}$, cf.~the discussion in Appendix~\ref{subsec:eigenspectrum_summary}.

\section{\label{sec:master}Derivation and the eigenspectrum of the master
equation}

\subsection{Derivation of Eq.~(\ref{eq:-diff-_equation_no_noise})}

The stochastic dynamics of the system is described by the probability
density $P_{t}(\theta)$ of being in the state $\ket{\psi(\theta)}$
at time $t$ for the stochastic variable $\theta$. The probability
of the system being in an interval of states $[\theta_{1},\theta_{2}]$
at time $t$, $P_{t}([\theta_{1};\theta_{2}])=\int_{\theta_{1}}^{\theta_{2}}d\theta P_{t}(\theta)$
obeys the evolution
\begin{equation}
\int_{\theta_{1}}^{\theta_{2}}d\theta P_{t+dt}(\theta)=\int_{\tilde{\theta}_{1}}^{\tilde{\theta}_{2}}d\tilde{\theta}P_{t}(\tilde{\theta})p_{0}(\tilde{\theta})+\Theta_{H}(\pi\in[\theta_{1};\theta_{2}])\int_{0}^{2\pi}d\tilde{\theta}p_{1}(\tilde{\theta})P_{t}(\tilde{\theta}),\label{eq:diff_eq_integral}
\end{equation}
where $\Theta_{H}(x\in[a;b])$ is 1 if $x\in[a,b]$ and $0$ otherwise,
and $p_{r}(\theta)$ are the probabilities of obtaining the readout
$r$ in a measurement,
\begin{equation}
p_{r=1}\equiv p_{1}=\alpha dt\sin^{2}\frac{\theta(t)}{2},\quad p_{r=0}\equiv p_{0}=1-p_{1}.
\end{equation}
The first term on the r.h.s. of Eq.~(\ref{eq:diff_eq_integral})
describes the change of $P_{t+dt}([\theta'_{1};\theta'_{2}])$ due
to the the smooth evolution under $r=0$ readout, while the second
term accounts for jumps to $\theta=\pi$ for the measurement outcome
$r=1$. The variables $\tilde{\theta}_{1,2}$ are defined via the
self-consistent condition $\tilde{\theta}_{1,2}-\Omega(\tilde{\theta}_{1,2})dt=\theta_{1,2}$,
where $\Omega(\theta)=2\Omega_{s}\left[1+\lambda\sin\theta\right]$,
cf.~Eq.~(3) in the manuscript. A differential equation for $P_{t}(\theta)$
is obtained by solving the self consistent equation to order $(dt)^{2}$,
differentiating Eq.~(\ref{eq:diff_eq_integral}) over $\theta_{2}$,
and retaining the terms of order $dt$. With the explicit expressions
for $p_{r}$ and $\Omega(\theta)$, we get Eq.~(\ref{eq:-diff-_equation_no_noise}).

The integro-differential master equation (\ref{eq:-diff-_equation_no_noise})
needs to be supplemented by the boundary conditions. They are simple:
\begin{equation}
P_{t}(\theta=0)=P_{t}(\theta=2\pi).
\end{equation}
In other words, $P_{t}(\theta)=P_{t}(\theta+2\pi)$. With this condition,
it is easy to check that the normalization by total probability $\int_{0}^{2\pi}d\theta P_{t}(\theta)=1$
is preserved by this equation. Finally, it is useful for solving to
eliminate the integral part of the equation. This is easily done,
as it only contributes at $\theta=\pi$. Therefore, Eq.~(\ref{eq:-diff-_equation_no_noise})
is equivalent to
\begin{equation}
\frac{dP_{t}(\theta)}{dt}=-\alpha\sin^{2}\frac{\theta}{2}P_{t}(\theta)+\partial_{\theta}\left((2\Omega_{s}+\frac{\alpha}{2}\sin\theta)P_{t}(\theta)\right)
\end{equation}
at $\theta\neq\pi$, supplemented with boundary condition
\begin{equation}
2\Omega_{s}\left(P_{t}(\pi+0)-P_{t}(\pi-0)\right)=-\alpha\int_{0}^{2\pi}d\theta\sin^{2}\frac{\theta}{2}P_{t}(\theta).\label{eq:bc_pi}
\end{equation}

\subsection{\label{subsec:Eigenmodes}Eigenmodes of Eq.~(\ref{eq:-diff-_equation_no_noise})}

We now derive the eigenmode solutions of Eq.~(\ref{eq:-diff-_equation_no_noise}),
i.e., find all the solutions of the form $P_{t}(\theta)=e^{2\Omega_{s}\gamma t}f_{\gamma}(\theta)$.
The reader may skip the details and look at the result in Appendix~\ref{subsec:eigenspectrum_summary}.

Before diving into the derivation, it is useful to analyze the expectations
from the solution. Due to the normalization condition, for every eigenmode
with $\gamma\neq0$, we should have $\int_{0}^{2\pi}d\theta f_{\gamma}(\theta)=0$.
Since $P_{t}(\theta)\geq0$, there can be no eigenmodes with $\gamma>0$.
Thus, only solutions with $\gamma\leq0$ are acceptable. Finally,
since the decay rate of clicks at $\theta_{+}$ is $p_{1}(\theta)/dt=\alpha\sin^{2}\frac{\theta_{+}}{2}$,
for $\lambda>1$ we expect an eigenmode with $2\Omega_{s}\gamma\geq-\alpha\sin^{2}\frac{\theta_{+}}{2}=-\alpha\tan^{2}\frac{\theta_{+}}{2}/\left(1+\tan^{2}\frac{\theta_{+}}{2}\right)=2\Omega_{s}\left(\sqrt{\lambda^{2}-1}-\lambda\right)$,
i.e., $\gamma\geq\sqrt{\lambda^{2}-1}-\lambda$. We also expect a
steady state solution with $\gamma=0$ to exist.

\subsubsection{The functional dependence}

Away from $\theta=\pi$, the equation for the eigenmodes is

\begin{equation}
-\alpha\sin^{2}\frac{\theta}{2}f_{\gamma}(\theta)+\partial_{\theta}\left((2\Omega_{s}+\frac{\alpha}{2}\sin\theta)f_{\gamma}(\theta)\right)=2\Omega_{s}\gamma f_{\gamma}(\theta).
\end{equation}
Equivalently,
\begin{equation}
(2\Omega_{s}+\frac{\alpha}{2}\sin\theta)f_{\gamma}'(\theta)=\left[\frac{\alpha}{2}(1-2\cos\theta)+2\Omega_{s}\gamma\right]f_{\gamma}(\theta).\label{eq:steady_state_equation}
\end{equation}
Note that for $\alpha\geq4\Omega_{s}$, at $\theta=\theta_{\pm}$,
the equation becomes singular as the factor multiplying the highest
(and only) derivative vanishes. These singular points require special
treatment. However, this simply means that $f_{\gamma}(\theta\in[-\pi;\theta_{+}])=0$
as this interval is inaccessible from the time evolution of any state
initially outside it, as shown in the analysis of the postselected
dynamics in the manuscript. We will come back to the issue of the
special points later.

Away from the special points, the equation admits an analytic solution,
which can be expressed in two alternative forms

\begin{equation}
f_{\gamma}(\theta)=\frac{C}{(1+\lambda\sin\theta)^{2}}\exp\left(2\frac{\lambda+\gamma}{\sqrt{1-\lambda^{2}}}\left[\arctan\frac{\lambda+\tan\frac{\theta}{2}}{\sqrt{1-\lambda^{2}}}-\frac{\pi}{2}\right]\right)=\frac{C}{(1+\lambda\sin\theta)^{2}}\left(\frac{\tan\frac{\theta}{2}-\tan\frac{\theta_{+}}{2}}{\tan\frac{\theta}{2}-\tan\frac{\theta_{-}}{2}}\right)^{\frac{\lambda+\gamma}{\sqrt{\lambda^{2}-1}}},\label{eq:eigenmode}
\end{equation}
where $\lambda=\alpha/(4\Omega_{s})$. The equivalence of the two
expressions follows from
\begin{equation}
\arctan x=\frac{1}{2i}\ln\frac{1+ix}{1-ix},
\end{equation}
so one can write
\[
\left(\frac{\tan\frac{\theta}{2}-\tan\frac{\theta_{+}}{2}}{\tan\frac{\theta}{2}-\tan\frac{\theta_{-}}{2}}\right)^{y}=\left(\frac{\tan\frac{\theta}{2}+\lambda-\sqrt{\lambda^{2}-1}}{\tan\frac{\theta}{2}+\lambda+\sqrt{\lambda^{2}-1}}\right)^{y}=\exp\left[2y\left(\arctan\frac{\lambda+\tan\frac{\theta}{2}}{\sqrt{1-\lambda^{2}}}-\frac{\pi}{2}\right)\right].
\]
Obviously, the first form in Eq.~(\ref{eq:eigenmode}) is more convenient
for $\lambda<1$, while the second one is the natural choice for $\lambda>1$.
However, for computational purposes one can use either. Note the singularities
at $\theta=\theta_{+}$ and $\theta=\theta_{-}$ for $\lambda>1$
(for $\lambda<1$, $\tan\frac{\theta_{\pm}}{2}-\tan\frac{\theta}{2}\neq0$
for any $\theta$ since $\tan\frac{\theta_{\pm}}{2}$ has a non-zero
imaginary part while $\tan\frac{\theta}{2}$ is a real function).
From the equality
\begin{equation}
1+\lambda\sin\theta=\frac{1+2\lambda\tan\frac{\theta}{2}+\tan^{2}\frac{\theta}{2}}{1+\tan^{2}\frac{\theta}{2}}=\frac{\left(\tan\frac{\theta}{2}-\tan\frac{\theta_{+}}{2}\right)\left(\tan\frac{\theta}{2}-\tan\frac{\theta_{-}}{2}\right)}{1+\tan^{2}\frac{\theta}{2}},
\end{equation}
we obtain the following behavior in proximity of the singularity points
at $\theta=\theta_{\pm}$:

\begin{equation}
f(\theta\approx\theta_{+})\sim\left(\tan\frac{\theta}{2}-\tan\frac{\theta_{+}}{2}\right)^{\frac{\lambda+\gamma}{\sqrt{\lambda^{2}-1}}-2},\label{eq:singular1}
\end{equation}

\begin{equation}
f(\theta\approx\theta_{-})\sim\left(\tan\frac{\theta}{2}-\tan\frac{\theta_{-}}{2}\right)^{-\frac{\lambda+\gamma}{\sqrt{\lambda^{2}-1}}-2}.\label{eq:singular2}
\end{equation}
The normalizability conditions
\begin{equation}
\abs{\int^{\theta_{+}}d\theta f(\theta)}<\infty,\quad\abs{\int^{\theta_{-}}d\theta f(\theta)}<\infty.
\end{equation}
for Eqs. (\ref{eq:singular1}, \ref{eq:singular2}) imply that, for
$\lambda>1$, one must have
\begin{align}
\frac{\lambda+\mathrm{Re}\,\gamma}{\sqrt{\lambda^{2}-1}} & >1\text{ (vicinity of \ensuremath{\theta_{+}})},\\
\frac{\lambda+\mathrm{Re}\,\gamma}{\sqrt{\lambda^{2}-1}} & <-1\text{ (vicinity of \ensuremath{\theta_{-}})}.
\end{align}
These two conditions are incompatible. The apparent contradiction
is resolved by choosing the normalization constant $C$ in Eq.~(\ref{eq:eigenmode})
independently on intervals $(-\pi;\theta_{-})$, $(\theta_{-};\theta_{+})$,
and $(\theta_{+};\pi)$. Then for normalizable solutions, $C=0$ either
on the first two intervals or on the last two intervals. We do not
investigate the case of $C\neq0$ on $(-\pi;\theta_{-})$, as these
solutions (if exist) describe quick escape from the interval and cannot
contribute if the system is initialized outside of it. Choosing $C=0$
in the first two intervals, the eigenmodes are normalizable, and one
recovers the expected property $f_{\gamma}(\theta\in[-\pi;\theta_{+}])=0$.

Putting these results together, we have the general expression for
the eigenmodes, which reads, for $\lambda>1$,
\begin{equation}
f_{\gamma}(\theta)=\begin{cases}
\frac{C\left(1+\tan^{2}\frac{\theta}{2}\right)^{2}}{\left(\tan\frac{\theta}{2}-\tan\frac{\theta_{+}}{2}\right)^{2}\left(\tan\frac{\theta}{2}-\tan\frac{\theta_{-}}{2}\right)^{2}}\left(\frac{\tan\frac{\theta}{2}-\tan\frac{\theta_{+}}{2}}{\tan\frac{\theta}{2}-\tan\frac{\theta_{-}}{2}}\right)^{\frac{\lambda+\gamma}{\sqrt{\lambda^{2}-1}}}, & \text{for }\theta\in[\theta_{+};\pi],\\
0, & \text{for }\theta\in[-\pi;\theta_{+}];
\end{cases}\label{eq:eigenmode_polytan}
\end{equation}
and, for $\lambda<1$,
\begin{equation}
f_{\gamma}(\theta)=\frac{C}{(1+\lambda\sin\theta)^{2}}\exp\left(2\frac{\lambda+\gamma}{\sqrt{1-\lambda^{2}}}\left[\arctan\frac{\lambda+\tan\frac{\theta}{2}}{\sqrt{1-\lambda^{2}}}-\frac{\pi}{2}\right]\right),\quad\text{for }\theta\in(-\pi;\pi).
\end{equation}

\subsubsection{\label{subsec:bc+norm_integrals}The boundary conditions and normalization}

The above does not present the final solution. We have two boundary
conditions and a normalization condition to satisfy. The first boundary
condition, $f_{\gamma}(\theta)=f_{\gamma}(\theta+2\pi)$ is satisfied
trivially.

The second boundary condition, which needs to be addressed further,
Eq.~(\ref{eq:bc_pi}), yields
\begin{equation}
2\Omega_{s}\left[f_{\gamma}(\pi-0)-f_{\gamma}(\pi+0)\right]=\alpha\int_{-\pi}^{\pi}d\theta\sin^{2}\frac{\theta}{2}f_{\gamma}(\theta).
\end{equation}
Note that this condition is independent of $C$, hence it determines
the spectrum of eigenmodes $\gamma$.

The normalization condition is
\begin{equation}
\int_{0}^{2\pi}d\theta f_{\gamma}(\theta)=\begin{cases}
1, & \text{for }\gamma=0,\\
0, & \text{for }\gamma\neq0.
\end{cases}
\end{equation}
This should be used to determine $C$ for the steady state ($\gamma=0$)
and should be satisfied automatically for all $\gamma\neq0$ eigenmodes.
The integrals can be calculated analytically

\begin{multline}
\int d\theta f(\theta)=2C\int dt\frac{1+t^{2}}{\left(t-\tan\frac{\theta_{+}}{2}\right)^{2}\left(t-\tan\frac{\theta_{-}}{2}\right)^{2}}\left(\frac{t-\tan\frac{\theta_{+}}{2}}{t-\tan\frac{\theta_{-}}{2}}\right)^{\frac{\lambda+\gamma}{\sqrt{\lambda^{2}-1}}}\\
=\frac{C}{\lambda+\gamma}\left(\frac{t-\tan\frac{\theta_{+}}{2}}{t-\tan\frac{\theta_{-}}{2}}\right)^{\frac{\lambda+\gamma}{\sqrt{\lambda^{2}-1}}}\left(1+\frac{\lambda\left[(2\lambda+\gamma)(1-2\gamma t)-\gamma t^{2}\right]}{\left(t-\tan\frac{\theta_{+}}{2}\right)\left(t-\tan\frac{\theta_{-}}{2}\right)\left(\gamma-\tan\frac{\theta_{+}}{2}\right)\left(\gamma-\tan\frac{\theta_{-}}{2}\right)}\right),
\end{multline}
where $t=\tan\frac{\theta}{2}$. Similarly,

\begin{multline}
\int d\theta\sin^{2}\frac{\theta}{2}f(\theta)=2C\int dt\frac{t^{2}}{\left(t-\tan\frac{\theta_{+}}{2}\right)^{2}\left(t-\tan\frac{\theta_{-}}{2}\right)^{2}}\left(\frac{t-\tan\frac{\theta_{+}}{2}}{t-\tan\frac{\theta_{-}}{2}}\right)^{\frac{\lambda+\gamma}{\sqrt{\lambda^{2}-1}}}\\
=\frac{C}{2(\lambda+\gamma)}\left(t^{2}+\frac{(\gamma t-1)^{2}}{\left(\gamma-\tan\frac{\theta_{+}}{2}\right)\left(\gamma-\tan\frac{\theta_{-}}{2}\right)}\right)\frac{1}{\left(t-\tan\frac{\theta_{+}}{2}\right)\left(t-\tan\frac{\theta_{-}}{2}\right)}\left(\frac{t-\tan\frac{\theta_{+}}{2}}{t-\tan\frac{\theta_{-}}{2}}\right)^{\frac{\lambda+\gamma}{\sqrt{\lambda^{2}-1}}}.
\end{multline}

\subsubsection{The solution for $\lambda>1$ and \textmd{\textup{$\theta\in[\theta_{+};\pi]$.}}}

Assuming $\frac{\lambda+\mathrm{Re}\,\gamma}{\sqrt{\lambda^{2}-1}}>1$,
the boundary condition yields,
\begin{equation}
2\Omega_{s}C=\alpha\int_{\theta_{+}}^{\pi}d\theta\sin^{2}\frac{\theta}{2}f_{\gamma}(\theta)=\frac{\alpha C(2\gamma^{2}+2\lambda\gamma+1)}{2(\lambda+\gamma)(\gamma^{2}+2\lambda\gamma+1)}\Longleftrightarrow C\gamma(\gamma^{2}+\gamma\lambda+1)=0.
\end{equation}
This fixes the possible $\gamma$, thus giving us three eigenmodes:
\begin{equation}
\gamma=0,\gamma=\frac{1}{2}\left(-\lambda\pm\sqrt{\lambda^{2}-4}\right).\label{eq:eigenvalues}
\end{equation}
The norm of the eigenmodes is then
\begin{equation}
\int_{\theta_{+}}^{\pi}d\theta f(\theta)=\frac{C}{\lambda+\gamma}\left(1-\frac{\gamma\lambda}{(\gamma^{2}+2\lambda\gamma+1)}\right)=\frac{C}{\lambda+\gamma}\frac{\gamma^{2}+\gamma\lambda+1}{\gamma^{2}+2\lambda\gamma+1},
\end{equation}
which implies that the eigenfunctions with $\gamma=\frac{1}{2}\left(-\lambda\pm\sqrt{\lambda^{2}-4}\right)$
integrate to 0 as expected. For the steady state, $\gamma=0$, the
normalisation condition $\int_{\theta_{+}}^{\pi}d\theta f(\theta)=1$
yields
\begin{equation}
C=\lambda,
\end{equation}

\begin{multline}
f_{\gamma=0}(\theta)=\frac{\lambda\left(1+\tan^{2}\frac{\theta}{2}\right)^{2}}{\left(\tan\frac{\theta}{2}-\tan\frac{\theta_{+}}{2}\right)^{2}\left(\tan\frac{\theta}{2}-\tan\frac{\theta_{-}}{2}\right)^{2}}\left(\frac{\tan\frac{\theta}{2}-\tan\frac{\theta_{+}}{2}}{\tan\frac{\theta}{2}-\tan\frac{\theta_{-}}{2}}\right)^{1/\cos\theta_{+}}\\
=\lambda\frac{\left(1+\tan^{2}\frac{\theta}{2}\right)^{2}}{\left(\tan^{2}\frac{\theta}{2}+2\lambda\tan\frac{\theta}{2}+1\right)^{2}}\left(\frac{\tan\frac{\theta}{2}+\lambda-\sqrt{\lambda^{2}-1}}{\tan\frac{\theta}{2}+\lambda+\sqrt{\lambda^{2}-1}}\right)^{\lambda/\sqrt{\lambda^{2}-1}}.\label{eq:steady_state_polytan2}
\end{multline}

Finally, we check the conditions in Eqs. (\ref{eq:singular1}, \ref{eq:singular2})
\begin{equation}
\frac{\lambda+\mathrm{Re}\,\gamma}{\sqrt{\lambda^{2}-1}}>1\Longleftrightarrow\mathrm{Re}\,\gamma>\sqrt{\lambda^{2}-1}-\lambda.
\end{equation}
Interestingly, the condition actually requires that \emph{all} the
eigenmodes decay not slower than the decay rate at $\theta=\theta_{+}$.
The steady state, $\gamma=0$, always satisfies the condition for
$\lambda\in(1;+\infty)$. For $\lambda\leq2$, $\mathrm{Re}\,\frac{1}{2}\left(-\lambda\pm\sqrt{\lambda^{2}-4}\right)=-\lambda/2$.
Then, the normalizability condition is equivalent to
\begin{equation}
\lambda>2\sqrt{\lambda^{2}-1}\Leftrightarrow3\lambda^{2}<4\Leftrightarrow\lambda<2/\sqrt{3}.
\end{equation}
For $\lambda>2$, the inequality $\frac{1}{2}\left(-\lambda-\sqrt{\lambda^{2}-4}\right)<\frac{1}{2}\left(-\lambda+\sqrt{\lambda^{2}-4}\right)<\sqrt{\lambda^{2}-1}-\lambda$
always holds. Therefore, the eigenmodes $\gamma=\frac{1}{2}\left(-\lambda\pm\sqrt{\lambda^{2}-4}\right)$
are only normalizable for $\lambda<2/\sqrt{3}$.

\subsubsection{The solution for $\lambda<1$.}

The same considerations hold for $0\leq\lambda<1$. In this range
of parameters, $\sqrt{\lambda^{2}-1}$ and $\tan\frac{\theta_{\pm}}{2}$
become complex. Therefore, it is more convenient to use $f_{\gamma}(\theta)$
in the first form in Eq.~(\ref{eq:eigenmode}). At the same time,
the integrals for the norm and the boundary condition at $\theta=\pi$
are more conveniently calculated using the expressions in Eq.~(\ref{eq:eigenmode_polytan})
(see Appendix~\ref{subsec:bc+norm_integrals}). Using these expressions
and the results of Appendix~\ref{subsec:bc+norm_integrals}, one
finds that the boundary condition at $\theta=\pi$ yields
\begin{equation}
\frac{C}{\lambda+\gamma}\frac{\gamma^{2}+\gamma\lambda+1}{\gamma^{2}+2\lambda\gamma+1}\left(1-\exp\left[-2\pi\frac{\lambda+\gamma}{\sqrt{1-\lambda^{2}}}\right]\right)=0.
\end{equation}
Therefore, on top of the three eigenmodes, $\gamma=0$ and $\gamma=\frac{1}{2}\left(-\lambda\pm\sqrt{\lambda^{2}-4}\right)$,
there is also an infinite sequence of eigenvalues given by
\begin{equation}
\gamma=-\lambda+im\sqrt{1-\lambda^{2}},\quad m\in\mathbb{Z}.
\end{equation}
Since there are no special points for $\lambda<1$, both this infinite
set of solutions and the three eigenmodes in Eq. (\ref{eq:eigenvalues})
correspond to valid normalizable eigenmodes.

For the steady state, the normalization $\int_{-\pi}^{\pi}d\theta f(\theta)=1$
yields
\begin{equation}
C=\frac{\lambda}{1-\exp\left(-\frac{2\pi\lambda}{\sqrt{1-\lambda^{2}}}\right)},
\end{equation}
\begin{equation}
f_{\gamma=0}(\theta)=\frac{\lambda}{1-\exp\left(-\frac{2\pi\lambda}{\sqrt{1-\lambda^{2}}}\right)}\frac{1}{(1+\lambda\sin\theta)^{2}}\exp\left[\frac{2\lambda}{\sqrt{1-\lambda^{2}}}\left(\arctan\frac{\lambda+\tan\frac{\theta}{2}}{\sqrt{1-\lambda^{2}}}-\frac{\pi}{2}\right)\right].
\end{equation}
For $\gamma\neq0$,
\begin{equation}
\int_{-\pi}^{\pi}d\theta f(\theta)=\frac{C}{\lambda+\gamma}\frac{\gamma^{2}+\gamma\lambda+1}{\gamma^{2}+2\lambda\gamma+1}\left(1-\exp\left[-2\pi\frac{\lambda+\gamma}{\sqrt{1-\lambda^{2}}}\right]\right),
\end{equation}
which vanishes due to the boundary condition above, as expected.

\subsection{Summary of eigenvalues and eigenmodes of Eq.~(\ref{eq:-diff-_equation_no_noise})\label{subsec:eigenspectrum_summary}}

Putting together the results from Appendix~\ref{subsec:Eigenmodes},
we can summarize the solutions of the master equation (\ref{eq:-diff-_equation_no_noise})
as follows. The steady state, with eigenvalue $\gamma=0$, is given
for $\lambda<1$ by

\begin{equation}
f_{0}(\theta)=\frac{\lambda}{1-\exp\left(-\frac{2\pi\lambda}{\sqrt{1-\lambda^{2}}}\right)}\frac{1}{(1+\lambda\sin\theta)^{2}}\exp\left[\frac{2\lambda}{\sqrt{1-\lambda^{2}}}\left(\arctan\frac{\lambda+\tan\frac{\theta}{2}}{\sqrt{1-\lambda^{2}}}-\frac{\pi}{2}\right)\right],
\end{equation}
while for $\lambda>1$, it reads
\begin{equation}
f_{0}(\theta)=\begin{cases}
\frac{\lambda}{(1+\lambda\sin\theta)^{2}}\left(\frac{\tan\frac{\theta}{2}+\lambda-\sqrt{\lambda^{2}-1}}{\tan\frac{\theta}{2}+\lambda+\sqrt{\lambda^{2}-1}}\right)^{\lambda/\sqrt{\lambda^{2}-1}}, & \text{for }\theta\in[\theta_{+};\pi],\\
0, & \text{for }\theta\in[-\pi;\theta_{+}],
\end{cases}
\end{equation}
with $\tan\frac{\theta_{\pm}}{2}=-\lambda\pm\sqrt{\lambda^{2}-1}$,
$\lambda=\alpha/(4\Omega_{s})$, and
\begin{equation}
\left(\frac{\tan\frac{\theta}{2}+\lambda-\sqrt{\lambda^{2}-1}}{\tan\frac{\theta}{2}+\lambda+\sqrt{\lambda^{2}-1}}\right)^{\lambda/\sqrt{\lambda^{2}-1}}=\exp\left[\frac{2\lambda}{\sqrt{1-\lambda^{2}}}\left(\arctan\frac{\lambda+\tan\frac{\theta}{2}}{\sqrt{1-\lambda^{2}}}-\frac{\pi}{2}\right)\right].
\end{equation}
These are the expressions used in the main text -- cf.~Eqs.~(\ref{eq:steady},
\ref{eq:steady-strong}) therein.

Several comments are in order. First, the expressions for both $\lambda<1$
and $\lambda>1$ are the same except for the normalization constant,
which exhibits an essential singularity at $\lambda=1-\epsilon$.
Second, exactly at the transition $\lambda=1$,
\begin{equation}
f_{0}(\theta)=\begin{cases}
\frac{1}{(1+\sin\theta)^{2}}\exp\left[-\frac{2}{1+\tan\frac{\theta}{2}}\right], & \theta\in[-\pi/2;\pi],\\
0, & \theta\in(-\pi;-\pi/2),
\end{cases}
\end{equation}
which displays an essential singularity at $\theta=-\pi/2$. Due to
the nature of this singularity, the transition at $\lambda=1$ appears
smooth when looking at the steady state probability density. In this
sense it could be regarded as a crossover.

The eigenmode spectrum consists of two sets of eigenvalues:
\begin{itemize}
\item $\gamma=-\lambda+im\sqrt{1-\lambda^{2}},\quad m\in\mathbb{Z}$ with
\begin{equation}
f_{\gamma}(\theta)=\frac{C}{(1+\lambda\sin\theta)^{2}}\exp\left(2im\left[\arctan\frac{\lambda+\tan\frac{\theta}{2}}{\sqrt{1-\lambda^{2}}}-\frac{\pi}{2}\right]\right).
\end{equation}
These eigenvalues exist for $\lambda\leq1$, become massively degenerate
at $\lambda=1$, and disappear at $\lambda>1$. This disappearance
coincides with the opening of the forbidden region $(-\pi;\theta_{+})$.
Note thus that the $\lambda=1$ transition, which appears as a crossover
in the steady state behavior, presents a drastic change in the eigenmode
spectrum.
\item $\gamma=\frac{1}{2}\left(-\lambda\pm\sqrt{\lambda^{2}-4}\right)$
with
\begin{equation}
f_{\gamma}(\theta)=\frac{C}{(1+\lambda\sin\theta)^{2}}\exp\left(\frac{\lambda\pm i\sqrt{4-\lambda^{2}}}{\sqrt{1-\lambda^{2}}}\left[\arctan\frac{\lambda+\tan\frac{\theta}{2}}{\sqrt{1-\lambda^{2}}}-\frac{\pi}{2}\right]\right),\text{ for }\lambda<1,
\end{equation}
\begin{equation}
f_{\gamma}(\theta\in[\theta_{+};\pi])=\frac{C}{(1+\lambda\sin\theta)^{2}}\left(\frac{\tan\frac{\theta}{2}-\tan\frac{\theta_{+}}{2}}{\tan\frac{\theta}{2}-\tan\frac{\theta_{-}}{2}}\right)^{\frac{\lambda\pm\sqrt{\lambda^{2}-4}}{2\sqrt{\lambda^{2}-1}}},\text{ for }\lambda>1.
\end{equation}
These eigenmodes disappear at $\lambda>2/\sqrt{3}$. This disappearance
coincides with the steady state starting diverging at $\theta=\theta_{+}+\epsilon$.
\end{itemize}
Finally, note that the eigenvalues $\gamma=\frac{1}{2}\left(-\lambda\pm\sqrt{\lambda^{2}-4}\right)$
for $\lambda<2$ correspond to solutions with an oscillatory behavior
superimposed with a decay in time, while for $\lambda>2$ they give
steadily decaying in time solutions. This transition is identified
in the manuscript in terms of the average qubit polarization, and
has been identified in the detector's signal \citep{Li2014a} as well.
Here it appears as a property of the eigenvalues spectrum. Curiously,
these eigenvalues are non-physical for $\lambda>2/\sqrt{3}$, which
is before the transition is reached. At the same time, Eq.~(\ref{eq:-diff-_equation_no_noise})
does know about these eigenvalues for any $\lambda$, as we show in
Appendix~\ref{subsec:average_polarization}.

\end{document}